# RISC microprocessor verification

[1] MITUL S. NAGAR, [2] HARESH A. SUTHAR , [3] CHINTAN PANCHAL

[1]M.E. Student, Department E&C Of Engineering, Parul Institute Of Engineering And Technology, Vadodara, Gujarat.
[2]H.O.D. , Department E&C Of Engineering, Parul Institute Of Technology, Vadodara, Gujarat.
[3]Project Manager, ASIC Division, einfochips, Ahmedabad, Gujarat.

*Mtul.nagar.08ec@gmail.com,*
*hareshsuthar@rediffmail.com,chintanpanchal@einfochips.com*

***ABSTRACT**:* Today's microprocessors have grown significantly in complexity and functionality. Most of today's processors provide at least three levels of memory hierarchy, are heavily pipelined, and support some sort of cache coherency protocol. These features are extremely complex and sophisticated, and present their own set of unique verification challenges. Verification is clearly not a point tool, but is part of a process that starts from initial product conception and is to some degrees complete when the product goes to market. Functional verification is necessary to verify the functionality at RTL level. Complex micro-processors like ARM are high performance, low cost and low power 32-bit RISC processors. In our paper complex microprocessor is ARM cortex M3, developed for the embedded applications having low interrupt latency, low gate count, 3- stage pipelining, branch prediction, THUMB and THUMB-2 instruction set. Functional verification is used to verify that the circuit full fills each abstract assertion under the implementation mapping. we explore several aspects of processor design, including caches, pipeline depth, ALUs, and bypass logic.The verification was done concurrently with the design implementation of the processor.

.

## I: INTRODUCTION

Verification is the act of reviewing, inspecting or testing, in order to establish and document that a product, service or system meets regulatory or technical standards. This is a complex task, and takes the majority of time and effort in most large electronic system design projects.

This is particularly important for complex integrated circuits, such as microprocessors, because such circuits are extremely costly and time-consuming to design and fabricate. The advancement of semiconductor technology has made it feasible to integrate more than ten million transistors on a single chip and to operate at a clock speed more than 300MHz. This design complexity has resulted in the verification challenge of microprocessor both in academia and industry.

Integrated circuit designs typically undergo a variety of design verification steps before the manufacturing process begins to ensure that the design meets all design goals. This evaluation process allows designers to identify design shortcomings without the cost and delay of actually building integrated circuit devices prior to performing any evaluation.

Verification can be done in three different Types. Functional, formal and analog are the types. RISC microprocessors are verified using functional verification

Functional verification, in electronic design automation, is the task of verifying the logic design confirms to specification. In everyday terms, functional verification attempts to answer the question "Does this proposed design do what is intended?"

Objective behind verification is to check functionality of microprocessor design. Properties need to verify are fetch instruction, decode instruction, memory access and instruction result

Section II depicted verification at different levels and discuss the verification based on accessibility of aspect of the design. Accessibility of aspect at distinguish each level in form of boxes like black, white and gray

Section III depicted the architecture specification and design specification of the Identified blocks core and NVIC with sub blocks in detail

Section IV and V depicted different components of verification are plan and environment respectively. These two chapters are finding answer of below questions which will help in development of verification component.
Generally Verification plan development is divided in two steps: What to verify and How to verify?
Step one: What to Verify?
List of clearly defined features-to-verify. This is called feature extraction phase.
Step two: How to Verify?
After defining what exactly need to be verified, define how to verify them.

## II: VERIFICATION BASICS
*A. verification levels*

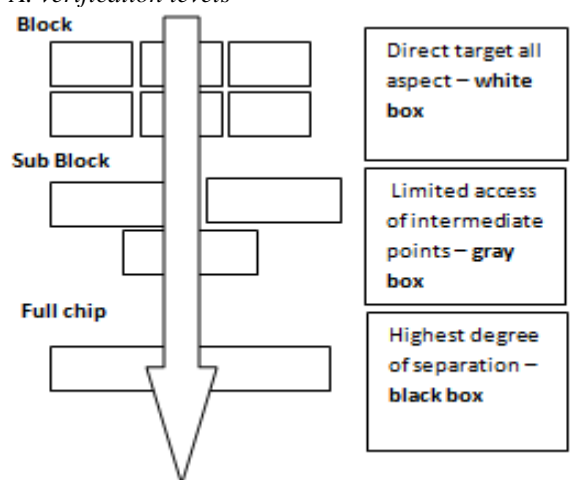

Figure 1 Verification level

During the design and verification process, verification will take place at a number of different levels. For the sake of this document, we will not be considering the effects of abstraction, or whether verification is performed in a top down or bottom up fashion. These factors have very little effect on the verification language choice. Three distinct phases will be considered, block level, sub-system and full chip. At the block level, the focus is on finding implementation problems within the block. At the sub-system level, it is the integration between the blocks that takes centre stage and at the full chip level, we are looking at end-to-end functionality [9].

The verification space can thus be considered as the cross of these two primary considerations, block size and complexity, and degree of separation between design and verification. If we analyze this grid, then not all points are equally useful. For example, black box verification techniques are not particularly well suited for doing block level implementation. The reason for this is somewhat pragmatic in that it becomes too difficult to reach all of the interesting corner cases when you know nothing about the implementation methodology. If this nformation is known, it becomes much easier to write very specific tests to target the corner case behaviour
.
*B. Libraries for verilog design environment*
In below figure 2 show the appropriate languages that fit well into these spaces. The sizes and positions of the blocks are meant to represent the relative capabilities and strengths of the solutions, but are all approximate and in some cases moved somewhat to allow all of the possibilities to be seen. it is assumed that whenever assertions are defined at the block level, they are left in during subsequent verification steps. So for example, the embedded assertion will still have value at the full chip verification stage, even though they may not have as high a value as when first created. This may not always be possible, especially if emulation becomes part of the system verification platform. While some of the assertion mechanisms are synthesizable

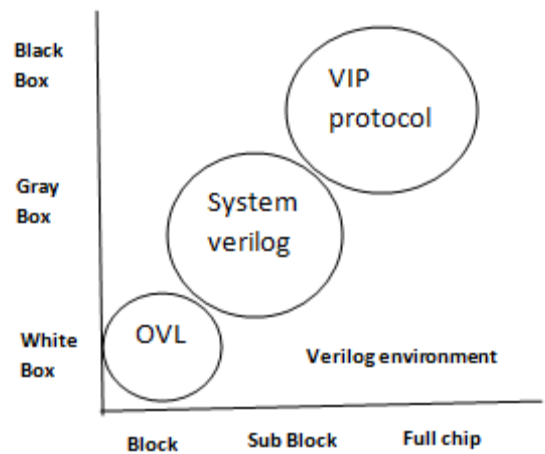

Figure 2 libraries verilog design environment

## III: ACHITECTURE OF ARM CORTEX M3
The Cortex-M3 is a low-power processor that features low gate count, low interrupt latency, and low-cost debug. It is intended for deeply embedded applications that require optimal interrupt response features.

The ARM cortex M3 can be defined in main parts as the core, Nested Vectored Interrupt Controller(NVIC), Memory Protection Unit(MPU), System Debug, Bus Mtrix, Wake up interrupt Controller

For sub block verification we have choose two blocks of ARM CORTEX m3. Core of the processor and NVIC is considered as synthesizable block for verificationCore of ARM CORTEX M3. *A low gate count processor core, with low latency interrupt processing[5].*

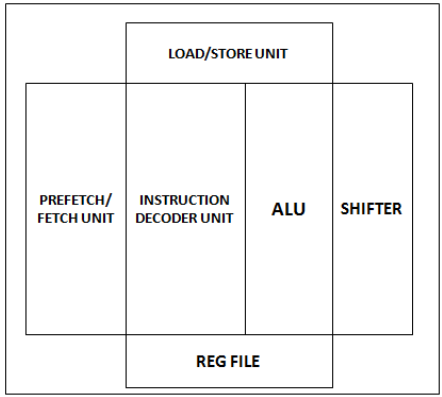

Figure 3 core internal diagram[1]

In figure 3 we have observe the internal blocks of core. These blocks like prefetch unit, instruction decode unit, ALU, shifter, REG file, Load/ Store unit, will become the sub block for verification.

*Nested Vectored Interrupt Controller*

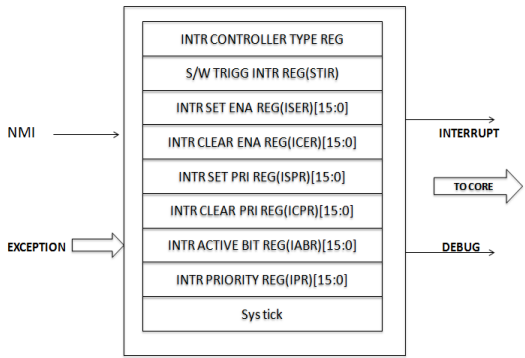

Figure 4 NVIC internal diagram

A *Nested Vectored Interrupt Controller* (NVIC) closely integrated with the processor core to achieve low latency interrupt processing. Figure 4 shows the internal structure of the NVIC block[5].

Today's contemporary designs are getting bigger and bigger in size, faster in speed and larger in complexity. This requires describing the design at the higher level of abstraction we have implement DUT in SYSTEM C. It is having some of advantages over other designing language like faster simulation, architecture exploration, hardware/ software co simulation[3][4].

### IV: VERIFICATION PLAN

The Verification Plan is the focal point for defining exactly what needs to be tested, and drives the coverage criteria. Success of a verification project relies heavily on the completeness and accurate implementation of a verification plan. A good plan contains detailed goals using measurable metrics, along with optimal resource usage and realistic schedule estimates. We will see the following steps for preparing verification plan see in figure 5.

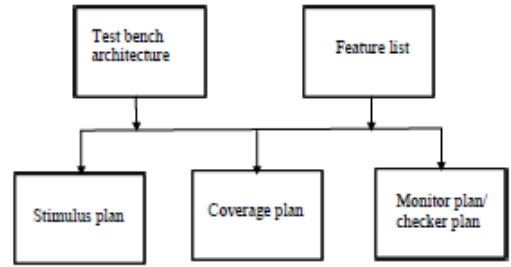

Figure 5 verification plan[12]

1) Overview: DUT, verification language and methodology
2) Feature extraction: unique name ID, description, expected result, priority
3) Verification environment
4) System verilog verification flow: block, sub block, full chip
5) Stimulus generation plan: unique name ID, stimulus for driver, configuration
6) Checker plan: checker unique name, unique feature ID, checker description
7) Coverage plan[12]

### V: VERIFICATION ENVIRONMENT

Verification environment consist of the test cases, generator, driver, monitor, chocker, scoreboard. This environment will write in system verilog. We need to understand mix language verification method because our DUT is written in System C and we will write our environment in system verilog.

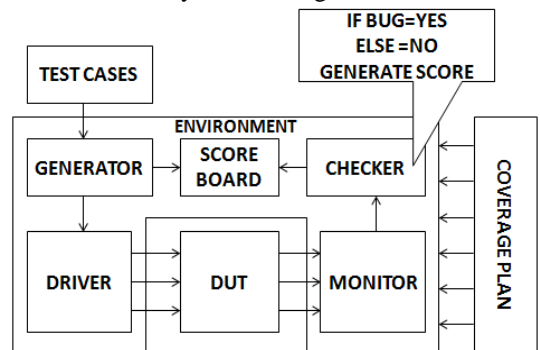

Figure 6 verification environment

**Stimulus:** In Verilog, the verification engineer is limited in how to model this stimulus because of the lack of high-level data structures. Typically, the verification engineer will create an array/memory to store the stimuli.

**Stimulus Generator** The generator component generates stimulus which are sent to DUT by driver. Stimulus generation is modeled to generate the stimulus based on the specification.
**Transactor:** It handles the DUT configuration operations. This layer also provides necessary information to coverage model about the stimulus generated.
**Driver** The drivers translate the operations produced by the generator into the actual inputs for the design under verification.
**Monitor** Monitor reports the protocol violation and identifies all the transactions. Monitors are two types, Passive and active. Passive monitors do not drive any signals. Active monitors can drive the DUT signals.
**Data Checker:** Checker converts the low level data to high level data and validated the data. This operation of converting low level data to high level data is called Unpacking which is reverse of packing operation.
**Scoreboard:** Scoreboard stores the expected DUT output.
**Coverage:** This component has all the coverage related to the functional coverage groups.
**Utilities:** Utilities are set of global tasks which are not related to any protocol.
**Environment:** Environment contains the instances of the entire verification component and Component connectivity is also done. Steps required for execution of each component is done in this[12].
**Tests** Tests contain the code to control the Test Bench features. Tests can communicate with all the Test Bench components.

*Testcases*

Different conditions are given directly or randomly to verify the design under test (DUT).

TABLE I.

| How To Verify | What To Observe | | |
|---|---|---|---|
| | signals | Bit | Description |
| ADD r1,r2,r3 [r1= r2+r3] | HTRANS [1:0] | "00" | IDEAL - prefetch unit is in ideal mode |
| | HSIZE [2:0] | "010" | WORD - 32 bit data is transferring |
| | HWRITE | "0" | READ - reading data from memory |
| | HBRUST [2:0] | "000" | TRANSFER SINGLE - single data is transferring |
| | I_CODE [31:0] | opcode | OPCODE - on I bus opcode transferring |
| | HREADY | "0" | TRANSFER EXTEND - show transfeing of data |
| | HRESP[1:0] | "00" | OKAY - if data successfully transfer give okay |

Sample test case for the ARM CORTEX M3 sub block verification

*Mix language modeling*

Current Design and Verification framework is very flexible and allows verification components to be written in many different ways, including different languages. it can be difficult to combine and reuse existing verification components coming from different sources and it often requires a considerable amount of effort to glue them together. [8].

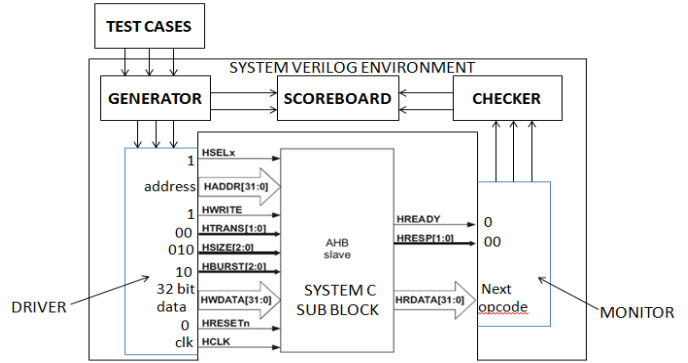

Figure 7 Test bench wrapping

We have DUT in system C and we will create the verification environment in system verilog we need to put efforts by using DPI direct port interface between DUT and testbench

## VI: CONCLUSION

With the help of sub block verification using system verilog we want to achieve good functional coverage and wants reduced the bottlenecks in the design of complex RISC processor
.
**REFRENCES**